# Contest between diverging k-vectors and astigmatism in an aberrated randomly scattered light

Abhijit Roy and Maruthi M. Brundavanam
Department of Physics, Indian Institute of Technology Kharagpur, W.B. 721302, India
Email: abhijitphy302@gmail.com

The impact of astigmatism on randomly scattered light, here speckles which are generated from a ground glass (GG) plate, is investigated by studying the change in the shape of the intensity correlation function (ICF) of the speckles for different amount of astigmatism introduced to it by rotating a lens at different angles. It is observed that the shape of the ICF, which reflects the average shape of the speckles, changes from circular to elliptical with the change of the tilt angle of the lens (θ), and the variation of length of two axes of the ICF with θ is completely different in nature. It is shown that these different natures are observed due to a contest between the diverging k-vectors of the speckles and the astigmatism introduced by the tilted lens system. It is also found that the diverging k-vectors resist the effect of astigmatism on the speckles, and after a certain θ, the impact of astigmatism becomes prominent. The contest between the diverging k-vectors and astigmatism in an aberrated speckle pattern are confirmed by further experimental studies and are discussed in detail. This study will surely be helpful in designing different imaging systems, where scattered light plays an important role.

The limitations in manufacturing different optical components such as lens, curved mirror etc. and its misalignment in an optical system leads to different kinds of aberration effects: spherical aberration, coma, astigmatism etc. [1]. The failure of marginal rays to follow the behavior of paraxial rays results in the spherical aberration, whereas an off-axis object leads to the effect of coma and astigmatism [1]. The presence of these aberrations in an imaging system, such as, camera, telescope etc. distorts the recorded images [1, 2]. Different efforts have been made to identify the sources of aberrations in an imaging system and to study its effect on the recorded images, and various techniques have also been developed to optimize these effects in the system [1]. The effect of astigmatism, due to a tilted surface of different curvatures, on various types of wavefronts have also been studied rigorously [1].

The propagation of a coherent beam of light through a random scattering medium or diffraction of scattered light from a lens results in the generation of a random granular intensity distribution, known as the speckles [3]. The study of speckles is useful in imaging and sensing through random scattering media [4-6], astronomy [7], non-invasive surface roughness measurement [8], different biomedical studies [9] etc. In most of these applications, the speckles are collected using a lens either far-field or camera lens, which exposes the speckles to different types of lens aberrations. So, investigating the impact of aberrations on speckles has become another area of interest because of the unavoidable presence of speckles in various coherent optical imaging systems. Besides, studying this effect can be useful to measure different kinds of primary and secondary aberrations of a camera lens [10].

It has been reported that speckle photography is vulnerable to optical aberrations, the presence of which leads to the loss of speckle correlation, and also results in the retrieval of incorrect information about dynamic objects [11]. In an imaging system involving dynamic objects, optical aberrations have been found to introduce extra displacement in the measurement leading to speckle decorrelation [12]. Besides, if the scattering spot size is smaller than the point spread function of the imaging system, the size and shape of the speckles depend on the aberrations, and the speckles become radially streaky in the case of spherical aberration [13]. A similar observation has also been made for extra-axial aberrations [14].

Different aberrations are found to be affecting the spatial statistics of speckles, and their effect on the autocorrelation, power spectral density and contrast of a speckle pattern, in the case of an imaging system, has been studied rigorously [15]. In a coherent system, the contrast of a fully developed speckle pattern has been found to be not impacted by optical aberrations [16]. On the other hand, in a partially coherent system, the speckle contrast is reported to be dependent on both the odd and even functional aberrations, whereas Gaussian laser speckles are influenced only by the even-functional aberrations [17]. It has been shown theoretically that if an aberrated lens is used to focus a Gaussian beam on a diffuser, the contrast of the generated speckles exhibits an oscillatory behavior with the increase in the strength of the lens introduced aberration for few particular combinations of aberration coefficient, beam parameter and the focal length of the lens [18]. The speckle statistics has also been observed to become non-circular Gaussian under the influence of optical aberrations, and it has been established that although the second order statistics of Gaussian laser speckles are affected by the even-functional aberrations, they remain unaffected by the odd-functional ones [19]. However, aberrations have been found to have no impact in retrieving orbital angular momentum through a scattering medium using higher-order Stoke correlation [20].

Although various efforts have been reported to understand the effect of optical aberrations on speckles and their impact, there is a lack of systematic study of these effects, especially in the case of astigmatism. In this work, we present a systematic study of the effect of aberrations due to a tilted lens system, which is a common source of aberrations, on laser speckles and we have shown that the tilted lens introduced astigmatism is resisted by the diverging k-vector of the speckles. The experimental observations are presented in detail.

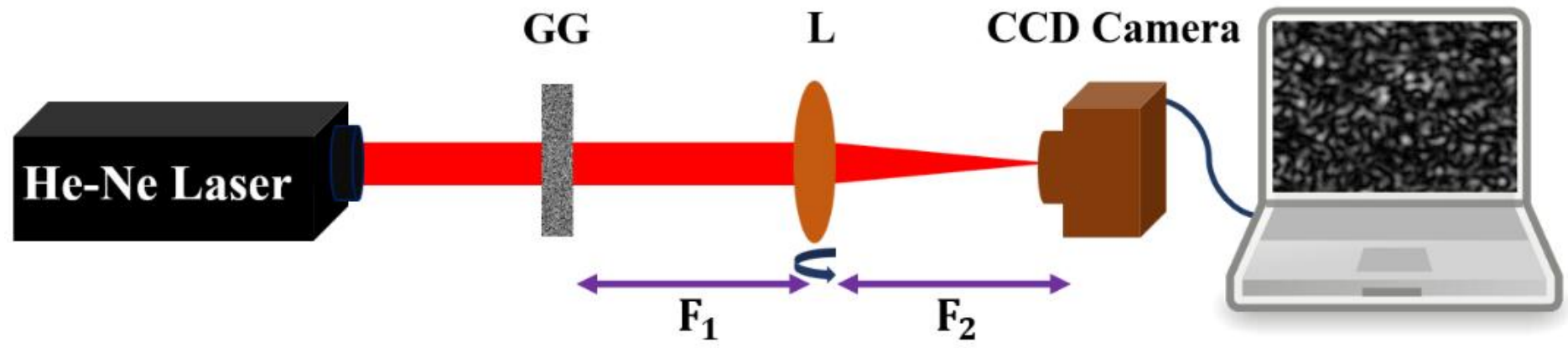

Fig. 1. Schematic of the experimental setup.

A schematic of the experimental setup to study the impact of astigmatism on laser speckles is presented in Fig. 1. A random scattering medium, here a ground glass (GG) plate, is illuminated with a beam of horizontally polarized light from a He-Ne laser source of 632.8 nm wavelength. The speckles generated from the GG plate are recorded by a charged coupled device, CCD camera. The GG plate and CCD camera are kept at a distance of $F_1$ and $F_2$ from the lens, L, as shown in the figure. The lens, L is placed on a rotational stage, and it is rotated at different angles (θ) w.r.t. the y-axis from $0^o$ to around $70^o$ in steps of $2^o$ to introduce astigmatism to the speckles, and for each θ, the aberrated speckles are recorded. The recorded speckle patterns for θ = $0^o$, $38^o$ and $60^o$ are presented in Figs. 2 (a-c), and it can be observed that the shape of the speckles are changing with θ.

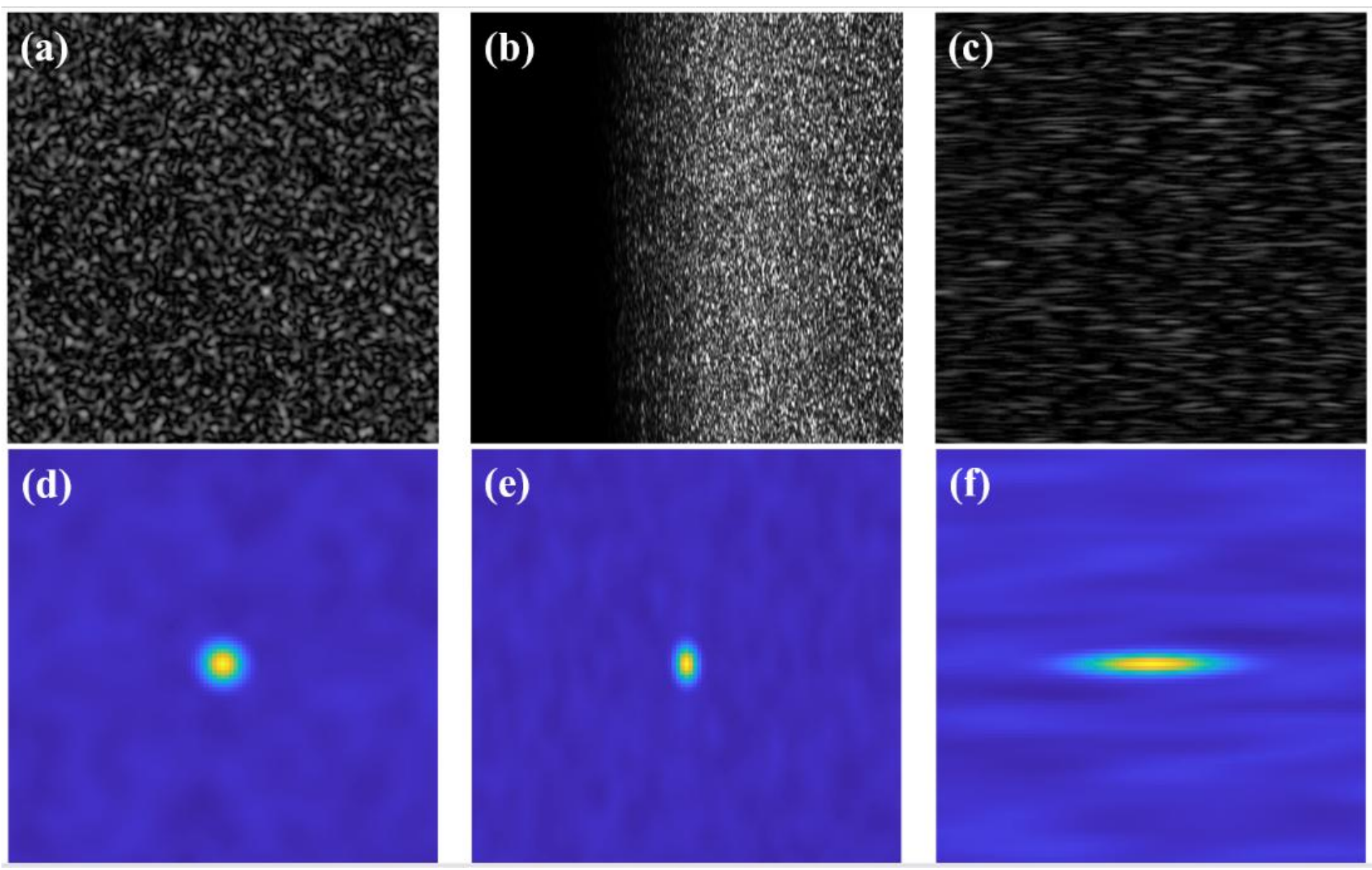

Fig. 2. For $F_1 = F_2 = 100$ mm, the recorded speckle patterns for θ = (a) $0^o$, (b) $38^o$ and (c) $60^o$, and the ICFs for θ = (d) $0^o$, (e) $30^o$, and (f) $60^o$.

The effect of the introduced astigmatism on the speckles is investigated by studying the two-point intensity correlation function (ICF) of the recorded speckles, following Ref. [21]. It is observed from the ICFs, presented for three values of θ in Figs. 2 (d-f), that the shape of the ICF changes with θ, i.e., with the amount of the introduced astigmatism. The variation in the shape of the ICFs with θ are characterized by studying the changes in the lengths of the major (along the horizontal direction) and minor (along the vertical direction) axes of the ICFs. These lengths are determined from the FWHM of the ICFs along their respective axis, and hence, can also be referred as the average speckle size along that direction. The change in the lengths of the major and minor axes of the ICFs with θ for $F_1 = F_2 = 100$ mm are presented in Fig. 3.

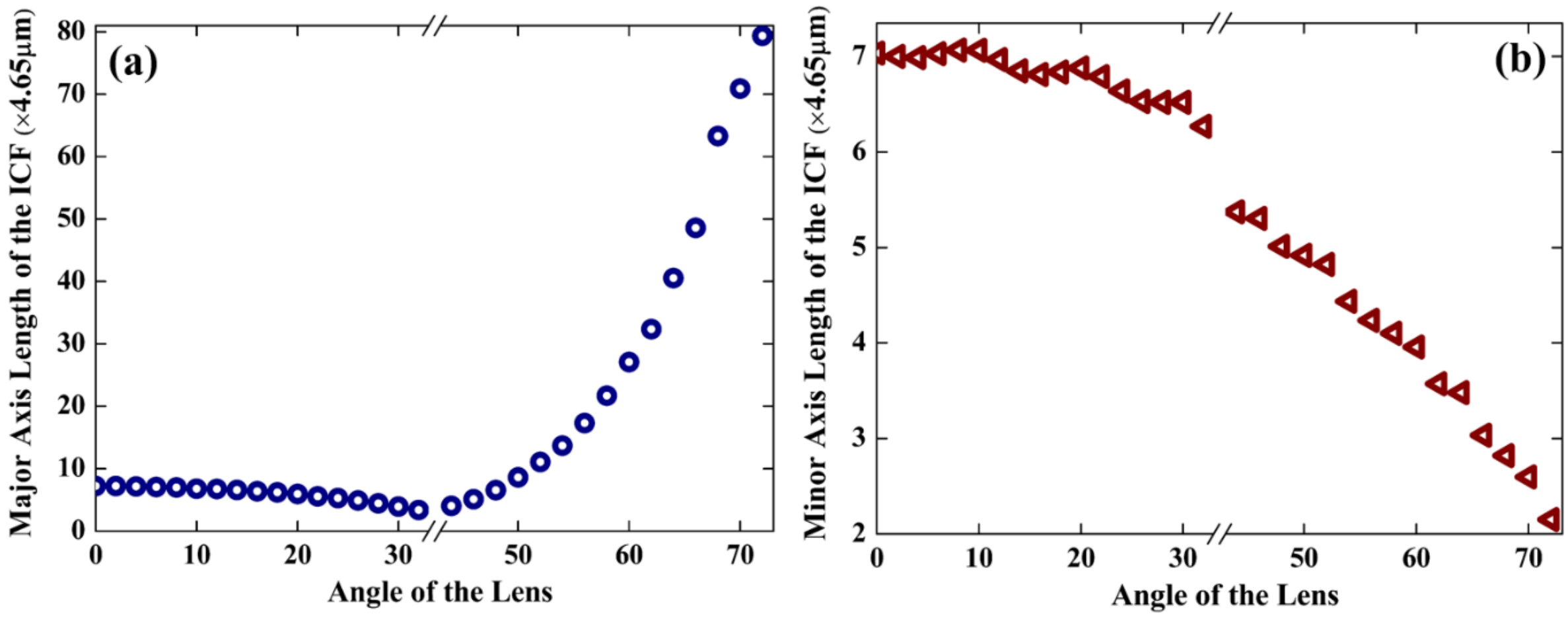

Fig. 3. Change in the lengths of the major and minor axes of the ICF with θ for $\mathrm{F_1 = F_2 = 100\ mm}$.

It can be observed from Fig. 3 that lengths of both the major and minor axes are same at $\theta = 0^0$, as expected in the case of a circular illumination on a GG plate. With the increase of θ, the length of the major axis is found to be decreasing initially and reaches to a minimum value, and then starts to increase rapidly (Fig. 3(a)). The observed discontinuity from $\theta = 32^o$ to $44^o$ is due to the fact that in this region, the speckle size along the horizontal direction reduces to less than two camera pixels, as evident from the plot, restricting to apply the intensity correlation technique due to the condition imposed by the Nyquist criteria. It is found that for $\theta = 32^o$ to $44^o$, the speckles get accumulated in one side (one example is shown in Fig. 2(b) for $\theta = 38^o$), and a maximum accumulation is observed at $\theta = 43.5^o$, which is denoted as $\theta_m$. On the other hand, the length of the minor axis of the ICF or the speckle size along the vertical direction decreases monotonically with the increase of θ (Fig. 3(b)), and at $\theta = 72^0$, this length is found to be close to two camera pixels, which indicates that beyond this angle, the intensity correlation technique cannot be employed to study this effect.

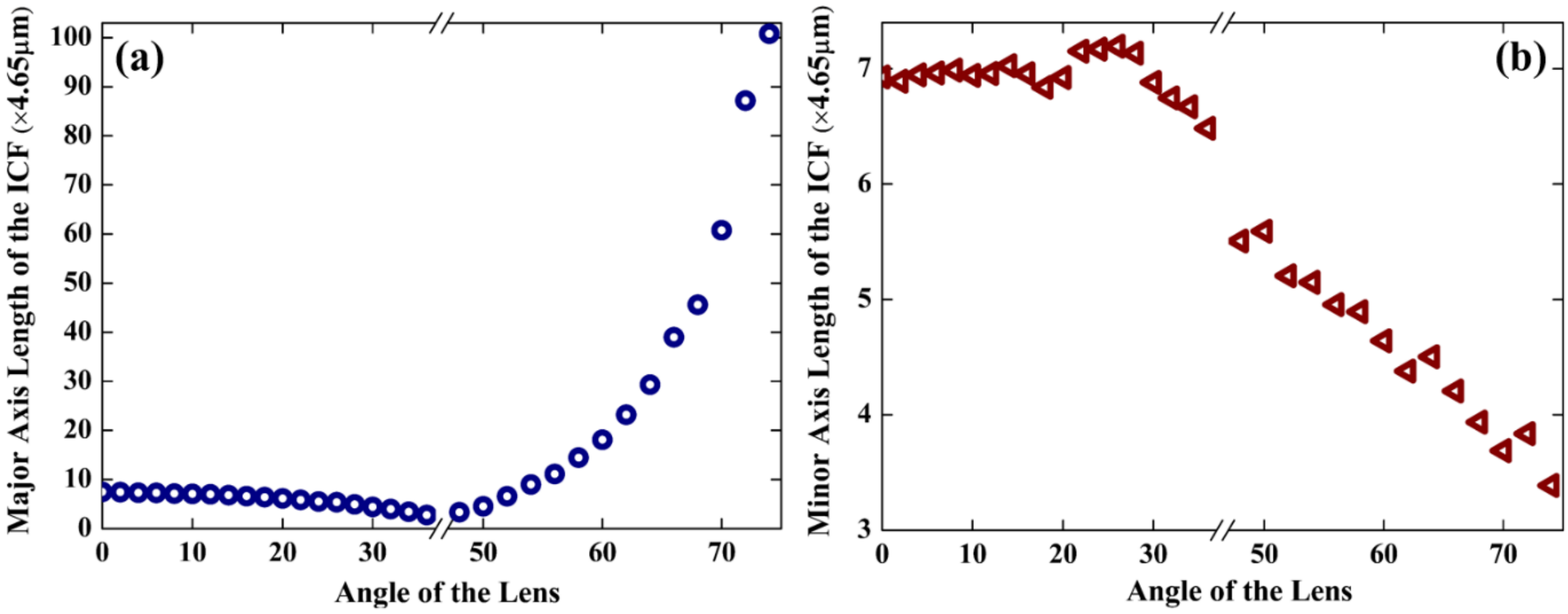

Fig. 4. Change in the lengths of the major and minor axes of the ICF with θ for $\mathrm{F_1 = 75\ mm}$ and $\mathrm{F_2 = 100\ mm}$.

It is observed that for $32^o < \theta < \theta_m$, with the increase of θ, the speckles start to shift and then accumulate in the horizontal direction and at $\theta = \theta_m$, the accumulation reaches to its maximum. The shift and accumulation of speckles are observed due to the refraction of speckles from the lens. In case of propagation of a plane wave through a similar tilted lens results in elongation of the beam along the horizontal direction due to the astigmatism induced by the tilted lens [1]. However, in the present case, shift and accumulation of speckles as well as decrease in speckle size along the horizontal direction are observed, which indicates that for $\theta < 44^o$, there is no effect of the lens induced astigmatism on the speckles. For $\theta > 44^o$, it can be found from Fig. 3 that the speckle size along the horizontal direction increases rapidly with the increase of θ, whereas the size along the vertical direction keeps on decreasing signalling that the speckles are elongated along the horizontal direction and squeezed along the vertical direction. This indicates that the observed changes in the speckle size for $\theta > 44^o$, may be due to the effect of the lens induced astigmatism. Hence, it can be concluded that the lens rotation does not immediately introduce astigmatism into the speckles. Considering the diverging k-vectors of speckles as one of the main differences between a plane wavefronts and speckles, it is assumed that the diverging k-vectors may be responsible for the behaviour observed in Fig. 3.

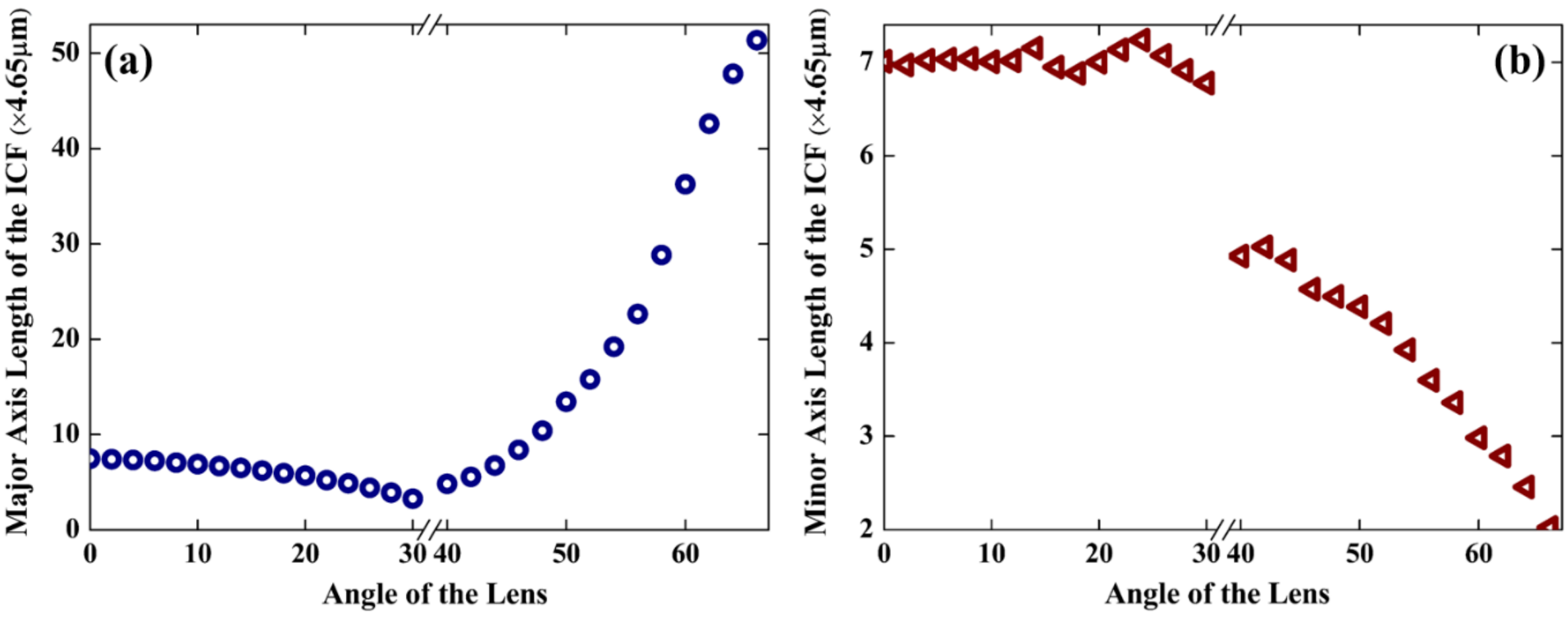

Fig. 5. Change in the lengths of the major and minor axes of the ICF with θ for $F_1 = 125$ mm and $F_2 = 100$ mm.

In order to confirm the effect of the diverging k-vectors, the following investigation is performed: the distance between the GG plate and the lens ($F_1$) is changed, and its impact on the variation of the major axis length of the ICF with θ is studied. As the change in $F_1$ results in changes in the divergence of k-vectors of the speckles illuminated on the lens, and the $\theta_m$ is expected to be changing in a systematic manner. The experiment, which is discussed earlier, is repeated by keeping the GG plate at a distance of 75 mm and 125 mm from the lens, while keeping the camera position unchanged, and the changes in the lengths of the major and minor axes of the ICFs with θ are presented in Figs. 4 and 5, respectively. The $\theta_m$ for $F_1 = 75$ mm and 125 mm is found be 43.5° and 35°, respectively, whereas for 100 mm, the $\theta_m$ is 38.5°, and this change as a function of $F_1$ is plotted in Fig. 6(a). It can be observed that with the increase in $F_1$, the $\theta_m$ decreases monotonically implying that as the divergence of the k-vectors decreases, the maximum accumulation of the speckles is observed at a smaller tilt angle. This study clearly shows that the variation of the major axis length of the ICF, presented in Fig. 3(a) for $\theta < \theta_m$, is due to the diverging k-vectors of the speckles passing through the lens. Besides, it can also be seen from Figs. 4 and 5 that the rate of increase in the major axis length for $\theta > \theta_m$ is different than that of in Fig. 3. This rate of change is studied by fitting it with an exponential function and the variation of the growth coefficient as a function of $F_1$ is plotted in Fig. 6(b), where it is found to be decreasing with the increase in $F_1$. This indicates that a speckle field with higher divergence resists more against the imposition of astigmatism, however, it also faces greater impact of astigmatism compared to a field with less divergence.

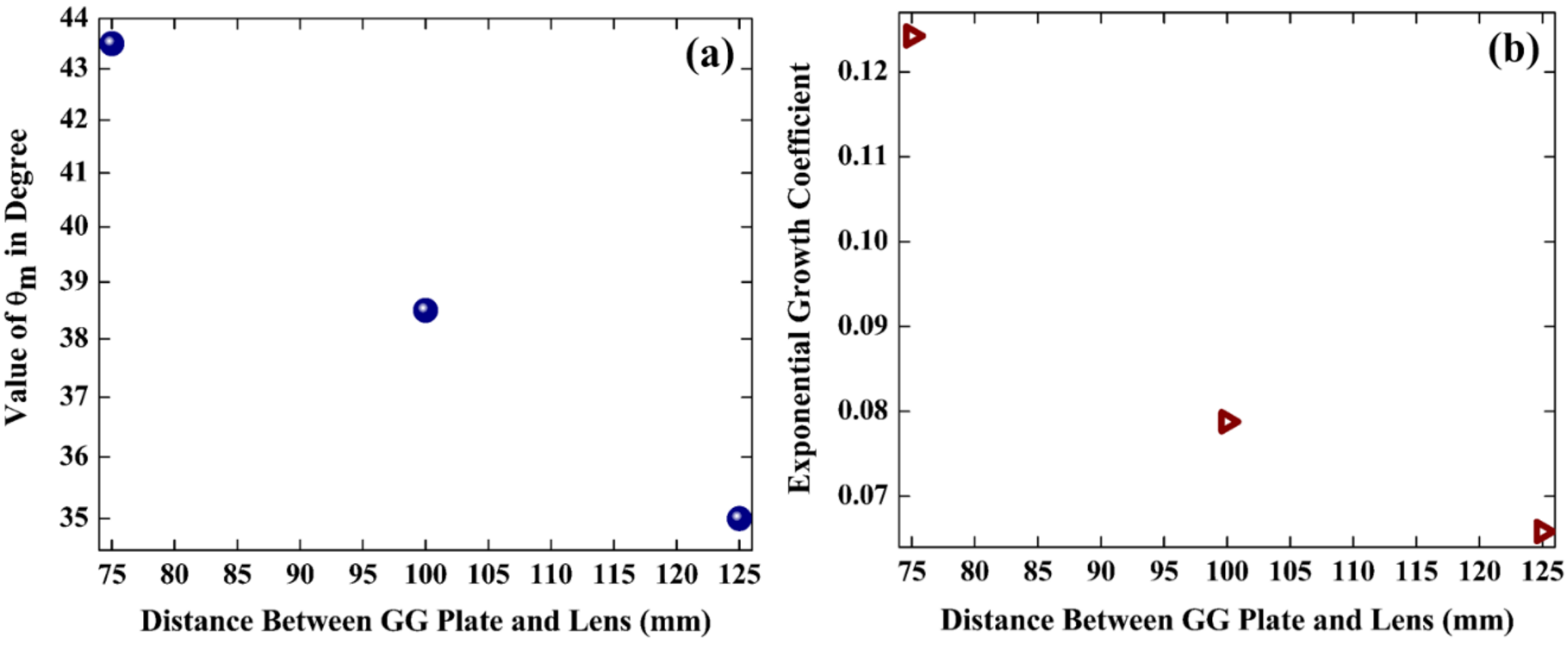

Fig. 6. Change in the value of $\theta_m$ and the exponential growth coefficient with $F_1$.

It is seen in Figs. 3, 4 and 5 that for $\theta > \theta_m$, the lengths of the major axes of the ICFs increase rapidly and that of the minor axes decrease, which is observed due to the astigmatism introduced by the tilted lens. This assumption is confirmed by replacing the bi-convex lens with a cylindrical lens of the same focal length, and by studying the changes in the ICFs of the recorded speckles for different orientations of the lens axis w.r.t. the z-axis. For the lens axis along the horizontal, diagonal and vertical directions, the recorded speckles and the ICFs of the speckles are presented in Figs. 7 (a - c) and 7 (d - f), respectively. It can be observed that the shape of the ICFs is elliptical, and their major axes are parallel to the lens axis, which indicates that the speckles are elongated along the axis of the cylindrical lens due to the astigmatism introduced by it. This study confirms that the effect observed in Fig. 3(a) for $\theta > \theta_m$ is due to the tilted lens introduced astigmatism. Hence, it can be concluded that the change of the speckle size along the horizontal direction after propagating through a tilted lens is due to the diverging k-vectors of the speckles for $\theta < \theta_m$ and the lens introduced astigmatism for $\theta > \theta_m$. Besides, when the distance between the GG plate and the lens is increased, the divergence of the k-vectors of the speckles reduces and the effect of astigmatism becomes prominent at smaller θ, which indicates that the diverging k-vectors compete with the lens introduced astigmatism.

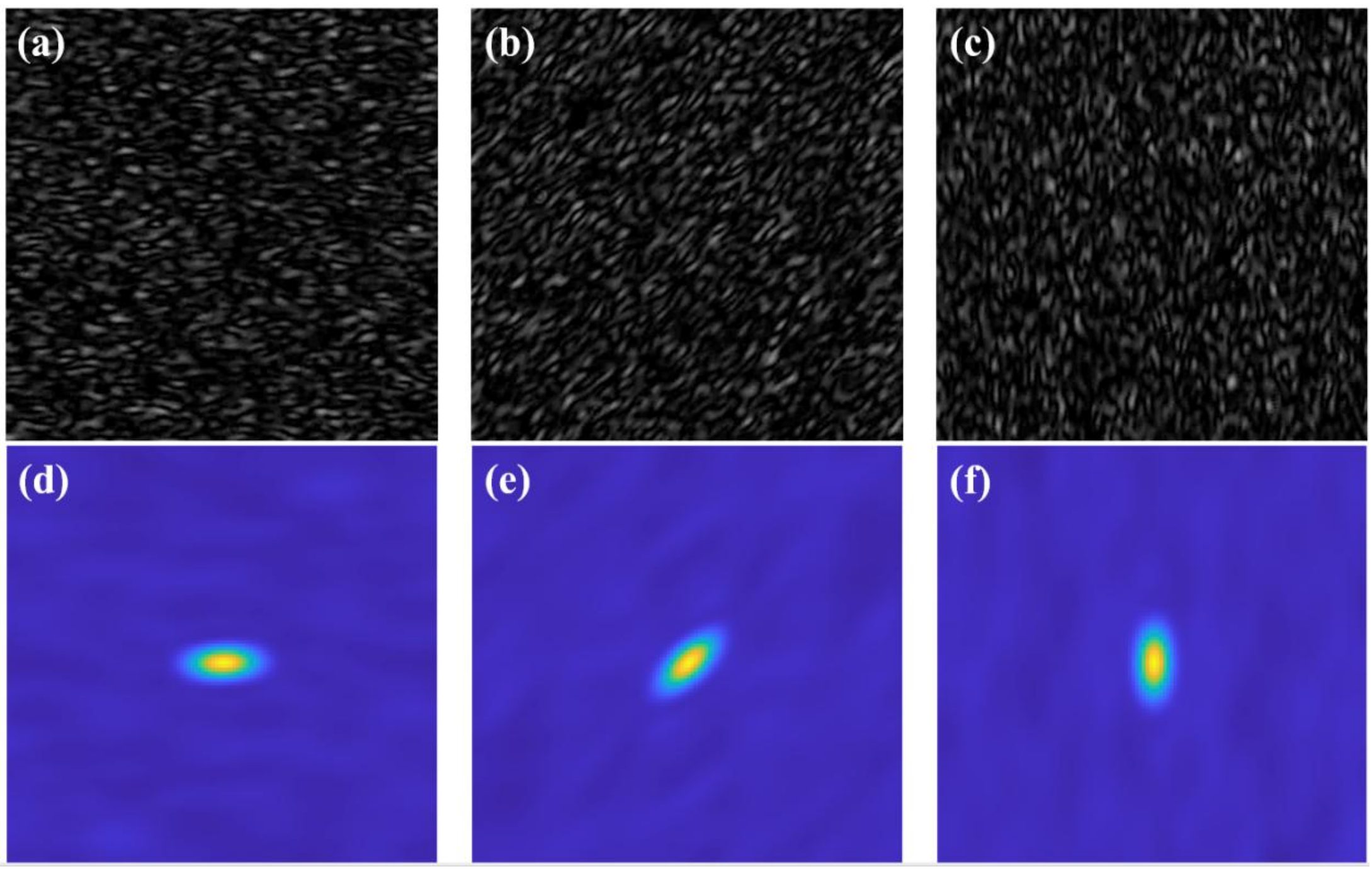

Fig. 7. The recorded speckle patterns (top row) and intensity correlation functions (bottom row) for the lens axis orientations: (a, d): $0^{o}$; (b, e): $45^{o}$; and (c, f) $90^{o}$.

In conclusion, we have investigated the impact of astigmatism, introduced by a tilted lens, on speckles. We have shown that the peculiar nature of the change in speckle size along the horizontal direction with the tilt angle of the lens is due to the diverging k-vectors of the speckles and the lens introduced astigmatism. It is also established that while propagating through a tilted lens, the diverging k-vectors of speckles compete with the lens introduced astigmatism and the impact of astigmatism is resisted by the diverging k-vectors. The astigmatic effect becomes prominent only after the effect of the k-vectors is nullified. The findings, which are reported here for speckles, may also be extended to any randomly scattered light and it will be useful in microscopic, biomedical and also in telescopic imaging system, where combination of lenses are used in a large scale, and an encounter with randomly scattered light having a large k-vectors is a common phenomenon. This study may also pave the way to generate an array of vortex beams using speckles exploiting a combination of two highly astigmatic lenses following the proposal of Beijersbergen et al. [22].